\documentstyle[12pt]{article}
\newcommand{\be}{\begin{equation}}
\newcommand{\ee}{\end{equation}}

\newcommand{\bea}{\begin{eqnarray}}
\newcommand{\eea}{\end{eqnarray}}

\newcommand{\p}{\partial}

\newcommand{\nn}{\nonumber \\}
\newcommand{\f}{\frac}
\newcommand{\w}{\wedge}

\begin{document}

\thispagestyle{empty}

\begin{flushright}
{\bf arXiv:0809.1756}
\end{flushright}
\begin{center} \noindent \Large \bf
More gravity solutions of AdS/CMT
\end{center}

\bigskip\bigskip\bigskip
\vskip 0.5cm
\begin{center}
{ \normalsize \bf  Shesansu Sekhar Pal}\\
\vskip 0.5cm

{ Barchana, Jajpur, 754081, Orissa, India \\
\vskip 0.5 cm

\sf shesansu${\frame{\shortstack{AT}}}$gmail.com}
\end{center}
\centerline{\bf \small Abstract}  We have generalized the gravity
solutions presented in arXiv:0808.1725 and arXiv:0808.3232 to
arbitrary but even space time dimensions with the scaling symmetry
$r \rightarrow \f{r}{\lambda} ,~~x_i \rightarrow \lambda^b x_i, ~~ t
\rightarrow \lambda^a t$. However, for $b=0$, we have the solution
in arbitrary space time dimension, not restricted to even
dimensional. The physical meaning of this particular choice of $b$ is 
that we can have a solution
with only temporal scale invariance. From the dual field theory
point of view, the BF bound and the unitarity bound for operators
dual to scalar field is discussed.

\newpage
 The usefulness of AdS/CFT correspondence \cite{jm},\cite{gkp}
and \cite{ew}, lies in solving the strongly coupled problems.
Moreover, it's important to  find the proper gravity solutions and
extracting  information from it by computing the correlation
functions.

In this context it has been proposed in \cite{klm} that we can have
a four dimensional gravitational description which exhibits the
Lifshitz-like fixed points with an exponent. This particular gravity
solution has been generalized in \cite{ssp} to two exponents and
more interestingly for a specific ratio of the exponents, there could be a
 possible dual field theory whose action is quadratic in fields. It could be
that we may be able to understand the strongly coupled physics of
some systems with quadratic in fields \cite{hls},\cite{gg} and
\cite{vbs}. In \cite{ssp}, the solution has been written in a better
coordinate system, in the sense to have only temporal scale
invariance along with the combination of spatial and temporal scale
invariance.

In a related context, several gravitational solutions has been
constructed in \cite{son} and  in \cite{bm} and  is being
embedded in string theory in \cite{hrr} -\cite{ssp1} that shows the
non-relativistic symmetry.

In all these  solutions the system exhibits the scaling symmetry and
we expect it to be preserved in both the bulk and boundary theories
\cite{jm}. In this paper, we would like to generalize the solutions
found in \cite{klm} and \cite{ssp} to arbitrary but even $d$
dimensions with both temporal scale invariance and a combination
of spatial and temporal scale invariance. Let us recall that these
solutions have been generated in four space time dimensions using
the electric type field for a two form field strength and a magnetic
type field for a three form field. Here, we also generalize that and
include the magnetic type field for the two form and electric type
field for the higher form field strength.

The form of the scaling symmetry that we are interested in is
\be\label{scaling_symmetry} r
\rightarrow \f{r}{\lambda} ,~~x_i \rightarrow \lambda^b x_i, ~~ t
\rightarrow \lambda^a t \ee

The action that we shall consider whose solution respects
eq(\ref{scaling_symmetry}) is \be \label{action}
S=\f{1}{2\kappa^2}\int d^d x \sqrt{-g}
(R-2\Lambda)-\f{1}{4\kappa^2}\int (F_2\w\star F_2+F_{d-1}\w\star
F_{d-1})-\f{c}{2\kappa^2}\int B_{d-2}\w F_2, \ee where
$F_{d-1}=dB_{d-2}$ and $c$ is the topological coupling. The equations
of motion  that follows from it are

\bea &&d\star F_2=-c F_{d-1},~~~d\star F_{d-1}=c F_2,\nn
R_{MN}&=&g_{MN}\bigg[\f{2\Lambda}{d-2}+\f{F^2_2}{4(2-d)}-\f{F^2_{d-1}}{2(d-1)!}
\bigg]+\f{1}{2}F_{MM_1}F_N{}^{M_1}\nn&+&\f{1}{2(d-2)!}F_{MM_1\cdots
M_{d-2}}F_N{}^{M_1\cdots M_{d-2}}\eea

The ansatz to the solution consistent with the scaling symmetry are
\bea\label{sol1}
ds^2&=&L^2[-r^{2a}dt^2+r^{2b}\sum^{d-2}_{i=1}\delta_{ij}dx^idx^j+\f{dr^2}{r^2}]\nn
F_2&=& \f{AL^2}{r^{1-a}}dr\w
dt,~~F_{d-1}=\f{BL^{d-1}}{r^{1-b(d-2)}}dr\w dx_1\w\cdots\w
dx_{d-2}\eea

It is trivial to see that the fluxes obey Bianchi identity and we
have considered electric field for the two form flux and magnetic
field for the $(d-1)$-form flux. The equations of motion associated
to the fluxes give \be c B L=A b (d-2),~~c A L= (-1)^{d-2} a B,\ee
and follows from the $F_2$ and $F_{d-1}$ fluxes, respectively.

In what follows, we shall stick to even $d$ dimensional space time. The
topological coupling that follows from the fluxes is \be c^2 L^2
=ab(d-2)\ee

From the metric \bea\label{ricci_component} R_{00}&=&a
r^{2a}[a+b(d-2)],~~R_{rr}=-\f{1}{r^2}[a^2+b^2(d-2)]\nn
R_{ij}&=&-br^{2b}[a+b(d-2)]\delta_{ij}\eea

From the equation of motion to metric \bea
R_{00}&=&-L^2r^{2a}\bigg[\f{2\Lambda}{d-2}-\f{3-d}{2-d}\f{A^2}{2}-\f{B^2}{2}\bigg]\nn
R_{ij}&=&L^2r^{2b}\bigg[\f{2\Lambda}{d-2}-\f{1}{2-d}\f{A^2}{2}\bigg]\nn
R_{rr}&=&\f{L^2}{r^{2}}\bigg[\f{2\Lambda}{d-2}-\f{3-d}{2-d}\f{A^2}{2}\bigg]\eea

Whose solution is \bea
A^2&=&\f{2a(a-b)}{L^2},~~B^2=\f{2b(a-b)}{L^2}(d-2)\nn\Lambda&=&-\f{a^2+ab(d-2)+b^2(d-2)^2}{L^2}\eea

It is interesting to note that for $ b=0$ and $c=0$, imply \be
A^2=\f{2a^2}{L^2},~~B^2=0,~~\Lambda=-\f{a^2}{L^2}\ee and this
solution is valid in any arbitrary space time dimension, not
necessarily restricted to even dimension. Moreover, in this case the
spatial directions $x,~y$ scales trivially, i.e. according to
eq(\ref{special_scaling}). This type of scale invariance suggests
that we  have a solution that possesses only the temporal scale
invariance.

Let us construct a possible action of the  dual field theory that
live on the boundary of eq(\ref{sol1}), consistent with the scaling
symmetry written in eq(\ref{scaling_symmetry}) \be S\sim \int dt
d^{d-2} x\bigg[ (\p^{m_1}_t\chi)^{\alpha}-K
(\p^{m_2}_x\chi)^{\beta}\bigg].\ee This form of the action gives the
restriction \be\alpha(am_1+h)=\beta(bm_2+h)=a+b(d-2),\ee where we
have assumed that the field, $\chi$, transforms under scale
transformation as $\chi\rightarrow \lambda^h \chi.$ If the field $\chi$
transforms trivially then the conditions are
$d-2=\f{m_2\beta}{m_1\alpha}[m_1\alpha-1]$ and
$\f{a}{b}=\f{m_2\beta}{m_1\alpha}$.

It is possible to have an action with  first order in time and
second order in space derivative and  quadratic in fields, $\chi$,
and  which imply $ a/b = 2$ and $d=4+2 \f{h}{b}$. For either
vanishing $h/b$ or any positive integer gives us the restriction
that the dimension of the corresponding spacetime should be even.
 Whereas if we want to
have an action with  second order in time and second order in space
derivative and  quadratic in fields, $\chi$, gives $a=b$ and
$d-1=4+2 \f{h}{b}$. From this it follows that  $d$ can be  an odd
integer, for either vanishing $h/b$ or any positive integer.

 Now let us restrict our attention to  a four dimensional theory with both
electric and magnetic type of fluxes for both $F_2$ and $F_3$ form
fluxes.

The ansatz to the fluxes are \bea F_2&=&A_1 L^2 r^{a-1}dr\w dt+A_2
L^2 r^{b-1}dr\w dx+A_3 L^2 r^{\widetilde{b}-1}dr\w dy,\nn F_3&=&B_1
L^3 r^{b+\widetilde{b}-1}dr\w dx\w dy+B_2 L^3
r^{a+\widetilde{b}-1}dr\w dt\w dy+B_3 L^3 r^{a+b-1}dr\w dt\w
dx\nn\eea

with the metric\be ds^2=L^2[-r^{2
a}dt^2+r^{2b}dx^2+r^{2\widetilde{b}}dy^2+\f{dr^2}{r^2}]\ee

In this case the scaling symmetry is \be\label{scaling_symmetry1} r
\rightarrow \f{r}{\lambda} ,~~x \rightarrow \lambda^b x, ~~y
\rightarrow \lambda^{\widetilde{b}} y,~~ t \rightarrow \lambda^a t
\ee

It is trivial to see that the field strengths obey Bianchi identity
and from the equation of motion that is \be d\star
F_2=-cF_3,~~d\star F_3=cF_2,\ee we get \be c^2
L^2=a(b+\widetilde{b}),~~c^2L^2=b(a+\widetilde{b}),~~c^2L^2=\widetilde{b}(a+b)\ee

The first relation comes from the coefficient that appear in the
fluxes $A_1$ and $B_1$, the second relation from $A_2$ and $B_2$ and the
third relation from $A_3$ and $B_3$, respectively. Along with
\be\f{A_1}{A_2}\bigg(\f{b+\widetilde{b}}{a+\widetilde{b}}\bigg)=\f{B_1}{B_2},~~
\f{A_1}{A_3}\bigg(\f{b+\widetilde{b}}{a+b}\bigg)=-\f{B_1}{B_3},~~
\f{A_2}{A_3}\bigg(\f{a+\widetilde{b}}{a+b}\bigg)=-\f{B_2}{B_3}\ee

The various components of Ricci tensor that follows from the metric are
\bea
R_{00}&=&a(a+b+\widetilde{b})r^{2a},~~R_{xx}=-b(a+b+\widetilde{b})r^{2b}\nn
R_{yy}&=&-\widetilde{b}(a+b+\widetilde{b})r^{2\widetilde{b}},~~
R_{rr}=-\f{1}{r^2}(a^2+b^2+\widetilde{b}^2)\eea

From the equation of motion to metric \bea R_{00}&=&\f{L^2}{2}
r^{2a}[\f{1}{2}(A^2_1+A^2_2+A^2_3)+B^2_1-2\Lambda]\nn
R_{xx}&=&\f{L^2}{2}
r^{2b}[\f{1}{2}(A^2_1+A^2_2-A^2_3)+B^2_2+2\Lambda]\nn
R_{yy}&=&\f{L^2}{2}
r^{2\widetilde{b}}[\f{1}{2}(A^2_1-A^2_2+A^2_3)+B^2_3+2\Lambda]\nn
R_{rr}&=&\f{L^2}{2r^2} [\f{1}{2}(-A^2_1+A^2_2+A^2_3)+2\Lambda]\eea

Finally comparing different components of Ricci tensor gives\bea
&&L^2\bigg[\f{1}{2}(A^2_1+A^2_2+A^2_3)+B^2_1-2\Lambda\bigg]=2a(a+b+\widetilde{b}),\nn
&&L^2\bigg[\f{1}{2}(A^2_1+A^2_2-A^2_3)+B^2_2+2\Lambda\bigg]=-2b(a+b+\widetilde{b}),\nn
&&L^2\bigg
[\f{1}{2}(A^2_1-A^2_2+A^2_3)+B^2_3+2\Lambda\bigg]=-2\widetilde{b}(a+b+\widetilde{b}),\nn
&&L^2\bigg[\f{1}{2}(-A^2_1+A^2_2+A^2_3)+2\Lambda\bigg]=-2(a^2+b^2+\widetilde{b}^2)\eea

It is interesting to note that there are seven parameters,
$A_i,~B_i$ for i=1, 2, 3 and $\Lambda$ and as many equations. It is
not that illuminating to find the most general solution. Instead, we
shall solve  these set of equations  in different cases.

For vanishing of all the flux coefficients i.e. $A_i=0$ and $B_i=0$
give the cosmological constant \be
\Lambda=-\f{3a^2}{L^2},~~ {\rm with}~~ a=b=\widetilde{b}\ee

 Let us consider a situation when one pair of coefficients,
$A_i,~B_i$, of fluxes is non-zero. The consistent solution i.e. real
flux is allowed when $A_1$ is non-zero and the rest coefficients are
zero and similarly for $B_2$ and $B_3$ separately.

For $A_1$ non-zero, the solution is \be
A_1=\f{2a^2}{L^2},~~\Lambda=-\f{a^2}{2L^2},~~b=0=\widetilde{b}\ee
and when $B_2$ is non-zero, the solution is \be
B_2=\f{4a^2}{L^2},~~\Lambda=-\f{2a^2}{L^2},~~b=0=\widetilde{b}\ee
and when $B_3$ is non-zero, the solution is \be
B_3=\f{4a^2}{L^2},~~\Lambda=-\f{2a^2}{L^2},~~b=0=\widetilde{b}\ee

It says that we can have a gravity solution by turning on either an
electric 2-form flux or a magnetic 3-form flux with a scaling
symmetry like \be\label{special_scaling} r \rightarrow \f{r}{\lambda} ,~~(x,~y) \rightarrow
 (x,~y), ~~ t \rightarrow \lambda^a t\ee

Instead of considering either a pure electric field or a magnetic field,
if we consider only $A_1$ and $B_1$ then the solution is presented
in \cite{ssp}. There is not any consistent solution for
considering $(A_2,~B_2)$ and $(A_3,~B_3)$ separately.

Let us take $(A_i,~B_i)$ and $(A_j,~B_j)$ for i $\neq$ j. In this
case the consistent solution is possible only for $(A_2,~B_2)$ and
$(A_3,~B_3)$ and the solution is \bea
A^2_2&=&A^2_3=\f{2b(a-b)}{L^2},~~B^2_2=B^2_3=\f{2(a^2-b^2)}{L^2},\nn
\Lambda&=&-\f{a^2+ab+b^2}{L^2},~~b=\widetilde{b}\eea

which respects the scaling symmetry as written in
eq(\ref{scaling_symmetry}). From this study it follows that we can
have a gravity solution with the scaling symmetry
eq(\ref{scaling_symmetry}), with the help of a magnetic two form flux
and an electric three form flux. Another thing to note that the ``exponents''
$b={\widetilde  b}$.

Now we would like to generalize it to arbitrary but even $d$
dimension. The ansatz to fluxes and geometry  are \bea\label{sol2}
F_2 &=&AL^2 r^{b-1}dr\w[dx_1+\cdots+dx_{d-2}]\nn
F_{d-1}&=&BL^{d-1}r^{a-1+b(d-3)}dr\w dt\w[dx_2\w\cdots\w
dx_{d-2}+\cdots+dx_1\w\cdots\w dx_{d-3}]\nn
ds^2&=&L^2[-r^{2a}dt^2+r^{2b}\sum^{d-2}_{i=1}\delta_{ij}dx^idx^j+\f{dr^2}{r^2}]\eea

The equation of motion to fluxes gives \be c^2 L^2=b[a+b(d-3)]\ee and
the equation of motion to metric gives\bea R_{00}&=&-L^2
r^{2a}[\f{2\Lambda}{d-2}-\f{A^2}{2}],~~ R_{ij}=L^2
r^{2b}[\f{2\Lambda}{d-2}+(d-3)\f{B^2}{2}]\delta_{ij}\nn
R_{rr}&=&\f{L^2}{r^2}[\f{2\Lambda}{d-2}+(d-3)\f{A^2}{2}]\eea

Solving for $A^2$, $B^2$ and $\Lambda$, using the components of
Ricci tensor as computed in eq(\ref{ricci_component}), results in
\bea
A^2&=&\f{2b(a-b)}{L^2},~~B^2=\f{2}{(d-3)L^2}[a^2-b^2(d-3)+(d-4)ab],\nn
\Lambda&=&-\f{(d-2)}{2L^2}[a^2+b^2+(d-3)ab] \eea

It is interesting to note that we can do a simple change of
coordinates \cite{ssp}, to bring the metric to the following form.
\be\label{metric_klm} ds^2=L^2[-\rho^{2\f{a}{b}}
dt^2+\rho^2\delta_{ij}dx^idx^j+\f{d\rho^2}{\rho^2}].\ee This form of
the metric coincides with the one written in \cite{klm} by defining
$z=a/b$ in four spacetime dimensions and this change of coordinates
makes sense only when $b \neq 0$. Which means in order to study only
temporal scale invariance, the coordinate system written in
eq(\ref{sol1}) and eq(\ref{sol2}) are better than
eq(\ref{metric_klm}).

\section{Field theory observable}

Generalizing the AdS/CFT prescription for the  case that we are
interested in makes us to  identify the operators dual
to bulk fields. In particular, we would like to identify the
dimension, $\Delta$, of an operator dual to scalar field of mass $m$
obeying the minimal scalar field equation. In $d$ space time dimension,
the dimension of the operator is
 \be
\Delta_{\pm}=\f{a+b(d-2)}{2}\pm\sqrt{\f{[a+b(d-2)]^2}{4}+m^2L^2}\ee

It follows that for the $\Delta_+$ branch there is a lower bound on
the dimension of the scalar operators that is $\f{a+b(d-2)}{2}$.

The requirement of the finiteness of the Euclidean action of the scalar field
imposes the restriction that
  if the mass of the scalar field obey \be m^2
L^2 > 1-\f{[a+b(d-2)]^2}{4},\ee then only $\Delta_+$ branch is
allowed, whereas if the mass stays \be -\f{[a+b(d-2)]^2}{4}< m^2 L^2
< 1-\f{[a+b(d-2)]^2}{4},\ee then both $\Delta_+$ and $\Delta_-$ branches are
allowed.

The analogue of Breitenlohner-Freedman bound \cite{bf} for this case
is \be (mL)^2 \geq -\bigg(\f{a+b(d-2)}{2}\bigg)^2 \ee and if the
mass stays below this bound then there is an instability in the
system.

The unitarity bound for the operators dual to scalar field depends
on the ratio of $a$ and $b$. For $a=2b$, the bound is \be \Delta >
\f{b(d-2)}{2}, \ee and for $a>2b$, the bound is \be \Delta >
\f{-a+b(d-2)}{2}, \ee whereas for $a < 2b$, the bound is \be \Delta
> \f{a+b(d-4)}{2}. \ee These bounds are computed following the
analogous prescription for the asymptotically AdS space time in
\cite{kw} i.e. demanding the positivity and finiteness of the
action of the scalar field in Euclidean space time.

The normalized solution to the Green's function of the scalar field
in the mass less limit in the $a=2b$ case \be
G(u,\overrightarrow{\mathbf{k}},\omega)=\f{\Gamma[\f{1}{2}+
\f{\overrightarrow{\mathbf{k}}^2}{4b\omega}+\f{d}{4}]}{\Gamma[\f{d}{2}]}e^{-\f{\omega}{2b}u^{2b}}
U[\f{\overrightarrow{\mathbf{k}}^2}{4b\omega}+\f{1}{2}-\f{d}{4},1-\f{d}{2},\f{\omega}{2b}u^{2b}],\ee
where $U(\alpha,\beta,z) $ is the confluent hypergeometric function
of the second kind.

Summarizing, we have presented  solutions in arbitrary even
dimensional spacetime dimensions with the scaling symmetry as
written in eq(\ref{scaling_symmetry}) as well as solution in any
arbitrary spacetime dimension with the scaling symmetry in
eq(\ref{special_scaling}). When the spacetime dimension is
restricted to even dimensional, we can have both the spatial and
temporal scale invariance and when $d$ is arbitrary, we have only
the temporal scale invariance. From the field theory point of view,
we have discussed the BF bound and the unitarity bound for operators
dual to scalar field.

\end{document}